\documentclass[preprint,aps,prb,10pt,showpacs,twocolumn,floatfix]{revtex4}
\usepackage{epsfig}
\begin{document}
\input epsf

\title{Computer simulation of
structural properties of dilute aqueous solutions of argon
at supercritical conditions.}
\author{V.~De~Grandis}
\author{P.~Gallo}\author{M.~Rovere}\email[Author to whom correspondence should be addressed: ]{rovere@fis.uniroma3.it}
\affiliation{Dipartimento di Fisica, 
Universit\`a ``Roma Tre'', \\ Istituto Nazionale per la Fisica della 
Materia, Unit\`a di Ricerca Roma Tre\\
Via della Vasca Navale 84, 00146 Roma, Italy.}

\begin{abstract}
Computer simulation studies of aqueous solutions of argon 
are performed from ambient to supercritical
conditions by using a recent polarizable potential model (BSV)
and the non polarizable simple point charge extended (SPC/E) model. 
At $T=673$~K we compare
the water-solute pair correlation
functions of the argon-water mixture with recent experimental
results obtained from neutron scattering experiments.
The comparison shows that the introduction of
the polarizable effects decreases the solute-water
repulsion and improves the agreement with the experiment
at supercritical conditions.
In particular
we find that the water-solute structure predicted by the
polarizable model is in good agreement
with the experiment.

\end{abstract}

\pacs{61.20.Ja, 61.20.-p, 61.25.-f}

\maketitle
\date{\today}

\section{ Introduction}

The problem of solubility of non polar solutes in water
has become of large interest for the applications in many
chemical processes in industrial technology and 
biochemistry.~\cite{shaw,debenedetti} 
At ambient conditions the solubility of non polar
gases in water is some order of magnitude lower than in other solvent
like liquid hydrocarbons, for this reason they are usually 
considered as hydrophobic solutes. The solubility however
increases steeply
above the water critical point, $T_c=647.13$~K and $P_c=220.55$~bar.     
Most applications involve the use 
of water at supercritical conditions, where it becomes a good
solvent for several non polar materials.~\cite{eckert}

In fact the curve of solubility of non polar gases at
increasing temperature shows a very peculiar
behaviour, starting from room temperature it decreases and reaches
a minimum, then it sharply increases toward
water supercritical temperature, where complete
miscibility is reached for some system.~\cite{crovetto}  It is also 
been observed that the solubility
of rare gases is larger for larger size of the 
solute.~\cite{guillot,lynden-bell}
The behaviour is
determined by the interplay
between energetic and entropic contributions to the
free energy of solvation. According to the simulation
work done by Guillot and Guissani~\cite{guillot} with the 
simple point charge extended (SPC/E) 
model~\cite{spce} for water the energetic term favors the 
solubility of larger solutes while the
entropic term depresses the solubility at increasing size.

The size of the solute seems also to be connected to the
different behaviour of the mixtures of water and noble gases
at high pressure at the gas-gas critical point.~\cite{schouten,schneider}

Structural and
computer simulation~\cite{guillot,tani} studies 
on aqueous solutions of rare gases
indicate that at ambient conditions
strong modifications of the water structure are
induced by the solutes with the formation of hydration cages
around the gas atom. 
Guillot and Guissani~\cite{guillot} found that at increasing temperature 
the solute tends to disrupt the cage. The entropic contribution is 
dominated by the reorientation
of the water molecules in the first shell of hydration, where only
the repulsive part of the water-solute interaction plays a role.

In this respect
the change in the structural properties of water 
approaching the supercritical states with the modification of the
hydrogen bond network~\cite{posto1,posto2,tromp} could 
be very relevant in determining
the strong increase of the solubility of apolar gases with temperature.
Recent neutron diffraction studies performed at supercritical
conditions in mixtures of rare gases at low concentration in water found
that the structure of water is modified and compressed by the presence
of the apolar solutes.~\cite{exper}

Simple pair potential used in computer simulation of water
based on parameters fitted to the properties at ambient
conditions like TIP4P~\cite{tip4p} and SPC/E~\cite{spce}
do not agree with the experimental results of neutron
diffraction in predicting
the structural changes in water at increasing 
temperatures.~\cite{chialvo1,chialvo2,kalinichev} 
Due to the interest in supercritical water several polarizable 
potentials have been recently developed with
the aim of reproducing the properties of water around and
above the critical point.~\cite{bsv,chialvo3,dang} 
According to a recent detailed analysis~\cite{jedlo2} 
performed by comparing the simulation 
with the more recent neutron diffraction data~\cite{bruni,soper,botti1}
the polarizable models give pair correlation functions which agree
with the experimental data in the supercritical region
much better than the nonpolarizable models.
On the other hand
the nonpolarizable models reproduce better the temperature
dependence of some thermodynamical quantities~\cite{jedlo2}
and the SPC/E potential for instance predicts a critical point
for water in agreement with the experiments~\cite{guissani},
while some polarizable models do not reproduce well the
coexistence curve.~\cite{kiyohara}
Polarizable models however are generally more accurate in describing the
modifications of the structure of water at increasing temperature
and pressure above the critical point.~\cite{jedlo2,jedlo1}

In this paper we present a comparison of the pair correlation functions
of mixtures of argon and water at supercritical conditions
obtained from computer simulation
with the recent experimental data quoted above.~\cite{exper}
In the numerical calculations water is simulated with
the SPC/E potential and 
the polarizable potential model proposed originally by
Ruocco and Sampoli~\cite{r+s1}
and further developed by Brodholt, Sampoli and Vallauri (BSV).~\cite{bsv}
This model is based on the TIP4P geometry with a polarizable
point dipole placed in the center of mass.
This potential has been checked in computer simulation of pure water
and found to produce pair correlation functions which are in good agreement 
with the experiment at
supercritical conditions.~\cite{jedlo2,jedlo1} 
In the next section we give some detail of our simulation. In Sec.~3
we present the results and compare them with recent
experimental data. Sec.~4 is devoted to conclusions.

\section{ Computer simulation of aqueous solutions}

We performed the computer simulations of a mixture of
$256$ water molecules and rare gas atoms 
at the concentration of one
atom for 40 molecules of solvent, the same used in the neutron
diffraction experiment quoted above. 
The interaction of 
the water molecules has been described by the polarizable
BSV model~\cite{bsv} and, for comparison, by the SPC/E model.~\cite{spce} 
In both the cases
the simulations have been
performed in the microcanonical ensemble with the
minimum image convention and a cut-off of the interactions
at half of the box length. 
The reaction field has been used
to take into account the long range part of the
electrostatic interactions.
The system with the SPC/E water has been simulated with the
DLPoly package in the
version 2.10, see reference~\cite{dlpoly} for further details.

The BSV potential has the same geometry as the TIP4P potential~\cite{tip4p}
with a polarizable dipole moment placed on the center of mass
of the molecule. The two positive charges, the parameters
of the Lennard-Jones potential on the oxygen and the position
of the negative charge are adjusted to reproduce some of the
properties of water at ambient conditions (see model 4 in ref.~\cite{bsv}).
The induced dipole $p_i=\alpha E_i$
is calculated from the local electric field $E_i$ with
an iterative procedure
by assuming an isotropic polarizability fixed to the value for water
molecules $\alpha=1.44$~\AA$^3$.~\cite{r+s1} The long range part of the 
electrostatic interactions
were taken into account by the reaction field method.~\cite{allen}
Details of the extension of the reaction field method
to include polarization effects can be found in references~\cite{r+s1,r+s2}. 

We used a time step of 
$10^{-15}$~s and the structural quantities have been
calculated by averaging over $2\cdot 10^5$ configurations. 
We carried out the simulations for
mixtures of water with Ar and Ne. Here we focus mostly on the results
of the simulation of water with argon atoms, for which
we can compare with recent experiments.~\cite{exper} 
The interaction of the water with the noble gas atoms has
been modeled with a Lennard-Jones potential between the oxygen
of the water molecules and the solute. The parameters of the 
Lennard-Jones potential  of argon are assumed as
$\epsilon_{Ar}/k_B=125$~K, $\sigma_{Ar}=3.415$~\AA\ and for neon
$\epsilon_{Ne}/k_B=18.56$~K and $\sigma_{Ne}=3.035$~\AA.~\cite{guillot} 
The parameters of
the oxygen-solute potential have been derived from the 
Lorentz-Berthelot rules.

We explored a range of temperature and
density from ambient conditions to the values in the supercritical
region where the experiment has been performed, as indicated
in Table~\ref{table1}. The calculations were performed below $T=673$
at the points which correspond
to the liquid branch of the coexistence curve of the
SPC/E model.~\cite{guissani} 
The point at $T=673$~K corresponds to the conditions of temperature
and density ($\approx 0.499$~g/cm$^3$) of the experiment.

\begin{table}
\caption{\label{table1}Thermodynamical points where the computer
simulations have been performed. The point at $T=673$~K corresponds
to temperature and density of the experiment.~\cite{exper}}
\begin{ruledtabular}
\begin{tabular}{cc}
$T$ (K) & $\rho$ (g/cm$^3$) \\
300  &  0.998 \\
373  &  0.949 \\
473  &  0.841 \\
673  &  0.499 \\
\end{tabular}
\end{ruledtabular}
\end{table}

\section{Solute-solvent structure}

\begin{figure}
\centering\epsfig{file=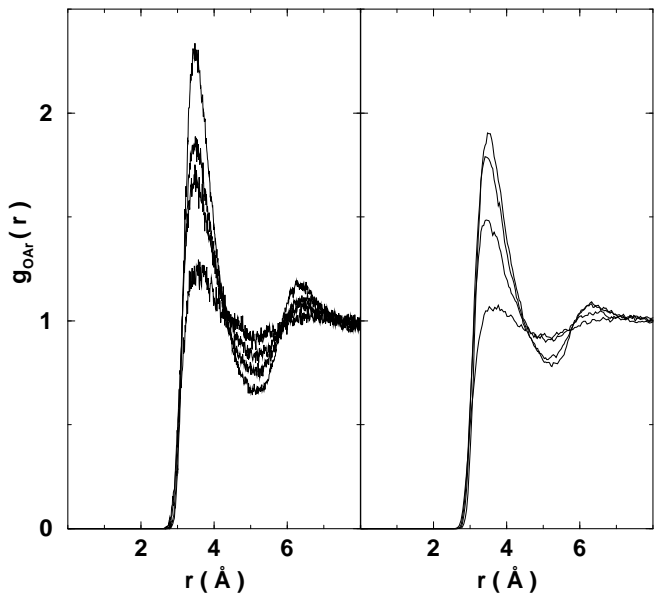,width=1\linewidth}
\caption{Oxygen-argon pair correlation functions on increasing
temperatures: comparison between the BSV model (on the left) and
the SPC/E model (on the right); temperatures 
from above are $T=300$~K, $T=373$~K, $T=473$~K, $T=673$~K.
The thermodynamical conditions are indicated in Table~\ref{table1}.
}
\protect\label{fig:goar}
\end{figure}

\begin{figure}
\centering\epsfig{file=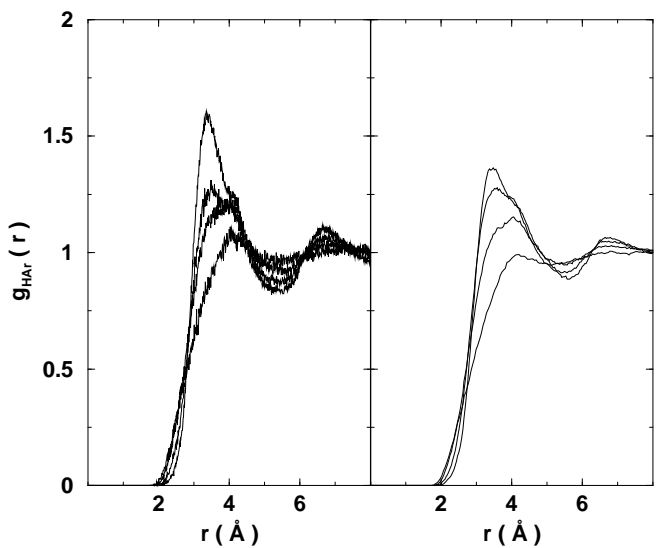,width=1\linewidth}
\caption{Hydrogen-argon pair correlation functions on increasing
temperatures: comparison between the BSV model (on the left) and
the SPC/E model (on the right); temperatures 
from above are $T=300$~K, $T=373$~K, $T=473$~K, $T=673$~K.
The thermodynamical conditions are indicated in Table~\ref{table1}.
}
\protect\label{fig:ghar}
\end{figure}

In Fig.~\ref{fig:goar} we report the pair correlation functions
argon-oxygen $g_{OAr}(r)$ at the
different thermodynamical points indicated in Table~\ref{table1}
obtained by the BSV and the SPC/E models.
At ambient conditions in both the models the $g_{OAr}(r)$ show
well defined first peak slightly different in the position due
to the different value of the Lennard-Jones parameter. The 
argon-hydrogen pair correlation functions $g_{HAr}(r)$ (Fig.~\ref{fig:ghar})
show a first peak at the same position as $g_{OAr}(r)$
indicating that the solute is located interstitially
in the hydrogen bond network equidistant on the average
from the oxygens and hydrogens.

The shoulder around $3.8$~\AA\ in the $g_{HAr}(r)$ indicates
however that some of the O-H bonds point radially toward the
second shell. Comparing the
positions of the second peak of the $g_{OAr}(r)$
with the one of the $g_{HAr}(r)$ it is evident that the
in the second shell the water molecules are preferentially
oriented with the oxygen atoms towards the solute and the
hydrogen atoms towards the bulk. 
Similar trends were already found by Guillot and Guissani on rare gases in
SPC/E water~\cite{guillot} and explained in terms of the formation
of an hydration cage around the solute (see~\cite{guillot} for more
details). 
We can conclude that the same effect is predicted by  
the polarizable BSV model at ambient conditions.

On increasing temperature the first and second peak of the $g_{OAr}(r)$
decrease and the second hydration shell tends to disappear.
In the $g_{HAr}(r)$ only the shoulder remains at the supercritical
conditions indicating that the solute tends to expel the water
molecules which are in its solvation shell. As it appears from
Fig.~\ref{fig:goar} and~\ref{fig:ghar} the BSV model confirms the
prediction of the SPC/E model and the microscopic scenario
described in ref.~\cite{guillot}, but it shows
a more drastic reduction of the peaks in going from ambient temperature 
to the point at $T=373$~K.

In the supercritical state at $T=673$~K the BSV potential,
which is more attractive with respect to the SPC/E model,
predicts a more structured $g_{OAr}(r)$ . We can
infer that in the process of expelling the solvent the depth
of the solute-solvent potential well plays an important role
in agreement with the conclusions of computer simulations~\cite{guillot}
and previous theoretical predictions~\cite{deben1,deben2}
on the relevance of the solute-solvent interaction
as compared with the solvent-solvent. 

\begin{figure}
\centering\epsfig{file=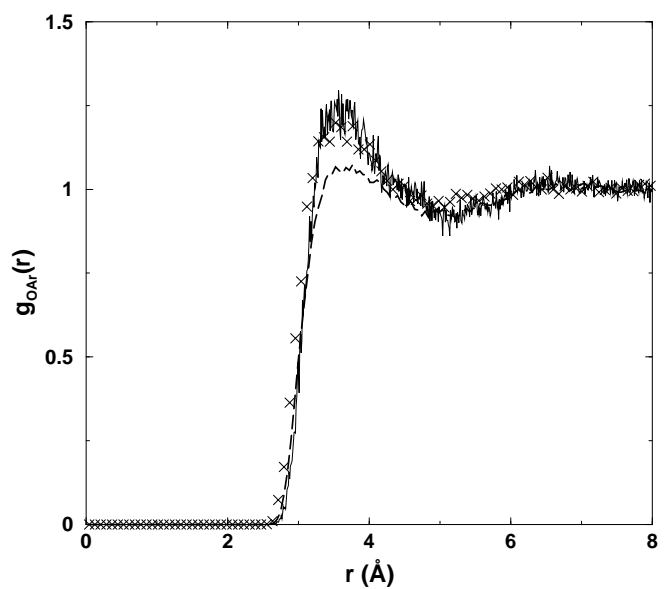,width=1\linewidth}
\caption{Oxygen-argon pair correlation function:
comparison between the BSV model (solid line), the SPC/E
model (long dashed line) and the experimental results~\cite{exper} (crosses)
at the thermodynamical conditions of the experiment ($T=673$~K) as
indicated in Table~\ref{table1}.
}
\protect\label{fig:oarvsexp}
\end{figure}

\begin{figure}
\centering\epsfig{file=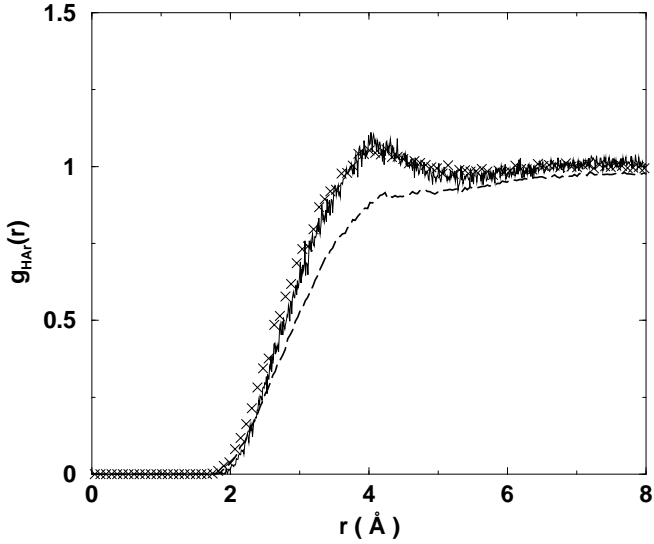,width=1\linewidth}
\caption{Hydrogen-argon pair correlation function:
comparison between the BSV model ( solid line), the SPC/E
model (long dashed line) and the experimental results~\cite{exper} (crosses)
at the thermodynamical conditions of the experiment ($T=673$~K) as
indicated in Table~\ref{table1}.
}
\protect\label{fig:harvsexp}
\end{figure}

In Fig.~\ref{fig:oarvsexp}-\ref{fig:harvsexp} 
the argon-water radial distribution functions obtained from our
simulation are compared with the experimental results~\cite{exper}.
In both $g_{OAr}(r)$ and $g_{HAr}(r)$ the use of BSV model improves
substantially the agreement with the experiment. The solute-solvent
effective interaction results to be more attractive in the
polarizable model with respect to SPC/E  as evident
from the comparison of the height of the first peak of the
radial distribution functions.

\begin{figure}
\centering\epsfig{file=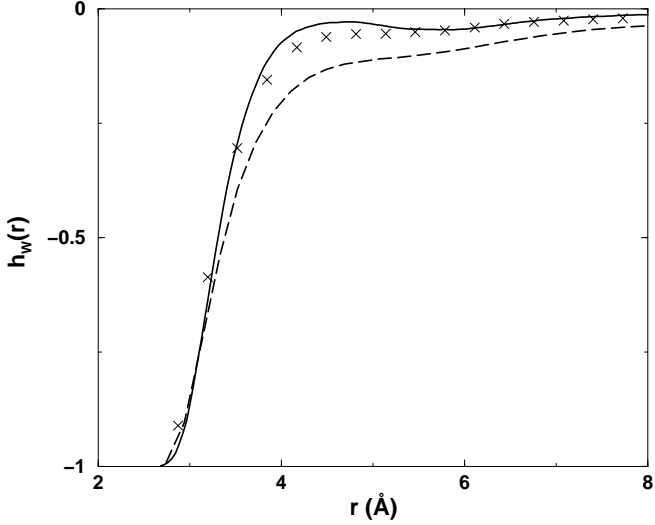,width=1\linewidth}
\caption{Deficit of water molecules around the solute, as
defined in the text (see Eq.~\ref{eq:deficit}), obtained
in the BSV (solid line) and in the SPC/E (long dashed line) models
compared with the same quantity calculated
from the experimental points (crosses). 
}
\protect\label{fig:deficit}
\end{figure}

This trend is confirmed by calculations of the coordination numbers
of the water molecules around the argon atoms. 
The deficit of water molecules around the solute can be estimated
by calculating the quantity ${h_w}(r)=\left[ n_{OAr}(r)/n_u(r) \right] -1$
where 
\begin{equation}
n_{OAr}(r) = 4 \pi \rho \int_0^r g_{OAr} (r') r'^2 dr' \label{eq:deficit}
\end{equation}
and the coordination number for a homogeneous fluid is given by
\begin{equation}
n_u(r) = \frac{4}{3} \pi \rho \left( r^3 - \sigma^3 \right)
\end{equation}
$\rho$ is the water density and $\sigma$ is the minimum approach
distance between argon and oxygen. 
The results are shown in Fig.~\ref{fig:deficit} compared with the quantity
$h_w(r)$ obtained from the experiments. The negative values indicate
a repulsive behaviour of the solute, but the repulsion is overestimated
by the SPC/E model particularly in the region of the first peak of
the $g_{OAr}(r)$.

\begin{figure}
\centering\epsfig{file=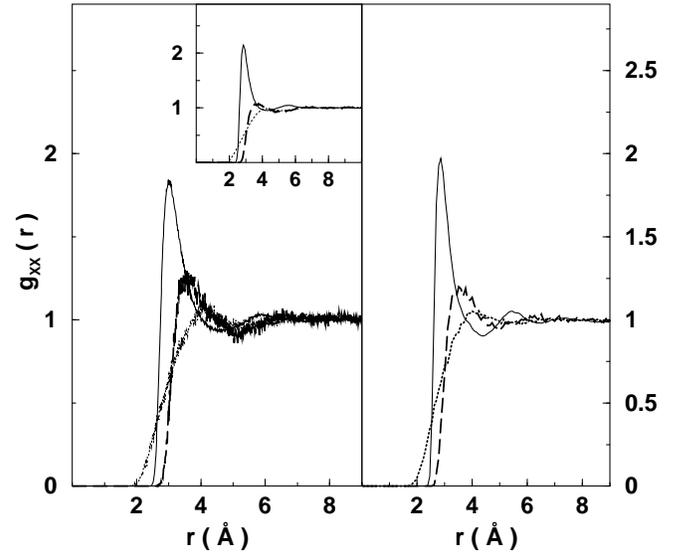,width=1\linewidth}
\caption{Radial distribution functions of the mixtures
at the supercritical conditions of the experiment 
at $T=673$~K (see Table~\ref{table1}): $g_{OO}(r)$
solid line, $g_{OAr}(r)$ long dashed line, $g_{HAr}(r)$
point line.
On the left results from the BSV model, on the right
results from the experiment~\cite{exper}, in the inset
results from the SPC/E model. 
}
\protect\label{fig:oooarhar}
\end{figure} 

In Fig.~\ref{fig:oooarhar} we report together the radial distribution
functions $g_{OO}(r)$, $g_{OAr}(r)$ and $g_{HAr}(r)$ for the 
water-argon mixture. On the right side there are the results of the
BSV, on the left the experimental findings and the inset shows the
SPC/E results. Both the models reproduce correctly how the
solute penetrates the shells of the water molecules. The $g_{OAr}(r)$ 
does not penetrate beyond the second shell, while the minimum
approach distance in the $g_{HAr}(r)$ is shorter than that observed in
the $g_{OO}(r)$. 

\begin{figure}
\centering\epsfig{file=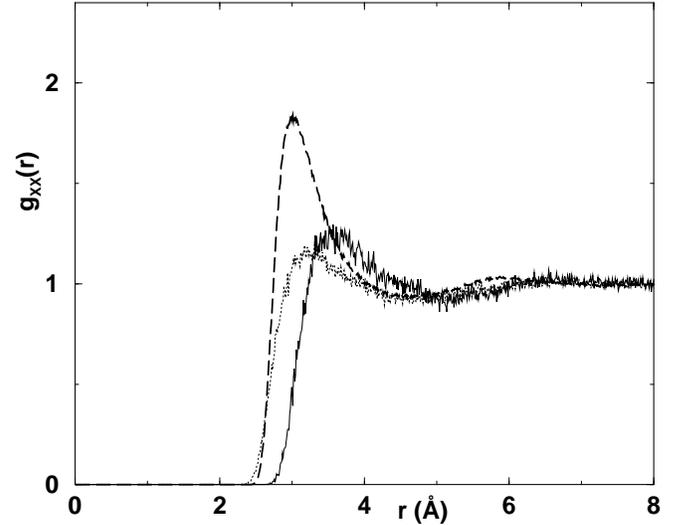,width=1\linewidth}
\caption{Radial distribution functions of the mixtures
from the BSV model
at $T=673$~K: $g_{OO}(r)$ (long dashed line), $g_{OAr}(r)$ (solid line), 
$g_{ONe}(r)$ (point line).
$g_{OO}(r)$ is extracted from the Ar-water simulation and it is shown
as representative of the position of the first O-O shell.
}
\protect\label{fig:oneoar}
\end{figure} 

\begin{figure}
\centering\epsfig{file=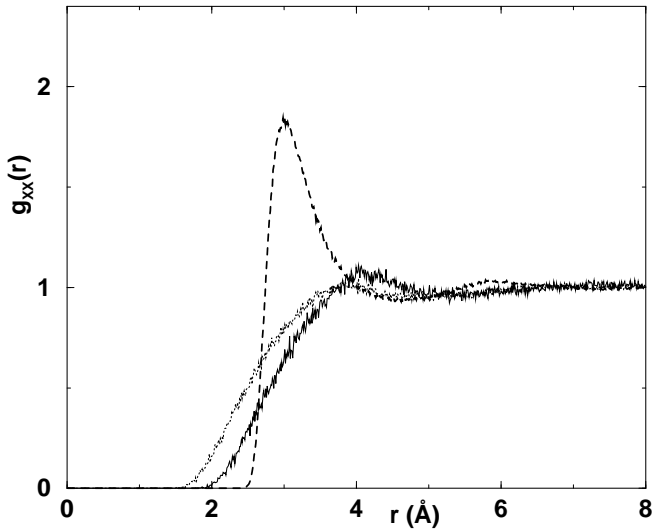,width=1\linewidth}
\caption{Radial distribution functions of the mixtures
from the BSV model
at $T=673$~K: $g_{OO}(r)$ (long dashed line), $g_{HAr}(r)$ (solid line), 
$g_{HNe}(r)$ (point line).
$g_{OO}(r)$ is extracted from the Ar-water simulation and it is shown
as representative of the position of the first O-O shell.
}
\protect\label{fig:harhne}
\end{figure}

In Fig.~\ref{fig:oneoar} we show the
solute-oxygen radial distribution functions of   
the water-argon and water-neon mixtures together with  
the $g_{OO}(r)$ of the argon-water mixture as obtained from the BSV model.
The calculations for the water-neon mixture were performed
at the same density and concentration of the water-argon
mixture, which correspond, as stated above, to the experimental
conditions. The $g_{OO}(r)$ of the two mixtures are very similar
and any of them could be chosen for the comparison.
It is evident the role of the size of the apolar solute. The
Ne atoms almost penetrate the first shell of water molecules. The
distance of minimum approach of Ne shifts to a lower value 
and roughly coincides with the minimum approach distance of the $g_{OO}(r)$.
The maximum of the $g_{ONe}(r)$ is at the same position of the
maximum of the $g_{OO}(r)$.
In Fig.~\ref{fig:harhne} it is shown that the $g_{HS}(r)$ 
of each solute have a minimum approach shorter than
that of the $g_{OS}(r)$.    
Our trend with the size of the solute 
is in agreement with the one found in the experiment.

\section{Conclusions}
We performed a computer simulation of a solution of argon atoms in water at 
the same conditions of a recent neutron diffraction
experiment~\cite{exper}. We compared the results obtained by using
the SPC/E model~\cite{spce} and the BSV polarizable potential.~\cite{bsv}
For both the
models we determined the solute-solvent radial distribution functions
from ambient to supercritical conditions. The local structure
of the solvent is perturbed by the presence of the solute with
the formation of a cage of hydration around the non polar atoms.
At increasing temperatures the solute tends to expel the water
which resides in the solvation shell. Both the models reproduce
this trend, at supercritical conditions however
the BSV potential predicts a more structured
$g_{OAr}(r)$ with respect to SPC/E as a consequence of a more
attractive solute-solvent interaction. The argon-water radial 
distribution functions are well reproduced at the experimental
conditions by the polarizable model. 
The SPC/E model 
overestimating the solute-solvent repulsion gives the effect
of collapsing the argon-water structure. 
The polarizable
model predicts much better the solvent-solute correlation
at supercritical conditions. 
This supports the idea that the BSV model works
quite well at high temperatures.~\cite{jedlo1}  
\section {acknowledgments}
The authors thanks M. A. Ricci for providing the experimental
results prior to publication and for useful stimulating discussions.
The authors also thank  M. Sampoli for providing the
computer code of the BSV Molecular Dynamics for bulk water.

\end{document}